\documentclass[%
aip,
rsi,
amsmath,assume,
reprint,%
]{revtex4-1}

\usepackage{babel}
\usepackage{graphicx}
\usepackage{dcolumn}
\usepackage{bm}

\usepackage[utf8]{inputenc}
\usepackage[T1]{fontenc}
\usepackage{mathptmx}
\usepackage{etoolbox}
\usepackage{enumerate}
\usepackage{color}
\usepackage{soul}
\usepackage{amsmath}

\makeatletter
\def\@email#1#2{%
 \endgroup
 \patchcmd{\titleblock@produce}
  {\frontmatter@RRAPformat}
  {\frontmatter@RRAPformat{\produce@RRAP{*#1\href{mailto:#2}{#2}}}\frontmatter@RRAPformat}
  {}{}
}%
\makeatother
\begin{document}

\preprint{AIP/123-QED}

\title[Prototyping Tools for CMOS Bioelectronic Sensors]{Wafer-Level Prototyping Tools for CMOS Bioelectronic Sensors}
\author{Advait Madhavan}
\altaffiliation{These authors contributed equally to this work.}
\affiliation{ 
Physical Measurement Laboratory, National Institute of Standards and Technology,\\Gaithersburg, MD 20899
}%

\author{Ruohong Shi}%
\altaffiliation{These authors contributed equally to this work.}
\affiliation{ 
Physical Measurement Laboratory, National Institute of Standards and Technology,\\Gaithersburg, MD 20899
}%
\affiliation{Johns Hopkins University, Baltimore, MD}

\author{Alokik Kanwal}
\affiliation{ 
Physical Measurement Laboratory, National Institute of Standards and Technology,\\Gaithersburg, MD 20899
}%

\author{Glenn Holland}
\affiliation{ 
Physical Measurement Laboratory, National Institute of Standards and Technology,\\Gaithersburg, MD 20899
}%

\author{Jacob M. Majikes}%
\affiliation{ 
Physical Measurement Laboratory, National Institute of Standards and Technology,\\Gaithersburg, MD 20899
}%
\affiliation{Theiss Research, LaJolla, CA}

\author{\\Paul N. Patrone}
\affiliation{ 
Information Technology Laboratory, National Institute of Standards and Technology,\\Gaithersburg, MD 20899
}%

\author{Anthony J. Kearsley}
\affiliation{ 
Information Technology Laboratory, National Institute of Standards and Technology,\\Gaithersburg, MD 20899
}%

\author{Arvind Balijepalli}
\affiliation{ 
Physical Measurement Laboratory, National Institute of Standards and Technology,\\Gaithersburg, MD 20899
}%
\email{arvind.balijepalli@nist.gov}


\begin{abstract}
Integrating biology with complementary metal-oxide-semiconductor (CMOS) sensors can enable highly parallel measurements with minimal parasitic effects, significantly enhancing sensitivity. However, realizing this potential often requires overcoming substantial barriers related to design, fabrication, and heterogeneous integration. In this context, we present a comprehensive suite of tools and methods designed for wafer-scale biosensor prototyping that is sensitive, highly parallelizable, and manufacturable. A central component of our approach is a new initiative that allows for open-source multi-project wafers (MPW), giving all participants access to the designs submitted by others. We demonstrate that this strategy not only promotes design reuse but also facilitates advanced back-end-of-line (BEOL) fabrication techniques, improving the manufacturability and process yield of CMOS biosensors. Developing CMOS-based biosensors also involves the challenge of heterogeneous integration, which includes external electrical, mechanical, and fluid layers. We demonstrate simple modular designs that enable such integration for sample delivery and signal readout. Finally, we showcase the effectiveness of our approach in measuring the hybridization of DNA molecules by focusing on data acquisition and machine learning (ML) methods that leverage the parallelism of the sensors to enable robust classification of desirable analyte interactions.
\end{abstract}

\maketitle

\section{\label{sec:intro}Introduction}



    


Rapid advances in both biomolecular technologies and complementary metal oxide semiconductor (CMOS) manufacturing have not yet translated into a broad ecosystem of commercial bioelectronic sensors that combine both approaches.\cite{liu_grand_2021, ostp_revitalizing_2022, semiconductor_research_corporation_2018_2018, forouhi_cmos_2019} This is not an accident. Widespread adoption is partly impeded by the complex requirements surrounding the development of integrated CMOS biosensors, which include access to robust prototype manufacturing, packaging, and testing tools. Smaller academic and commercial groups must overcome significant economic, supply chain, and technical challenges to access custom silicon devices produced by semiconductor manufacturers. Therefore, sustained research into manufacturable practices can improve the availability and use of electronic biosensors. Here, we show one approach to utilizing wafer-scale processing within a multi-facility multi-project wafer (MPW) framework to address the above core challenges.

In the long-term effort to develop integrated CMOS biosensors, there have been relatively few successful examples. These products include next-generation sequencing platforms that use millions of ion-sensitive field-effect transistors (ISFETs) to detect small changes in pH upon base incorporation.\cite{rothberg_integrated_2011} Commercial DNA sequencing now incorporates biological nanopores into individually addressable arrays with electronic readout and application-specific integrated circuits (ASICs) for on-device processing in a compact form factor.\cite{jain_oxford_2016} Additionally, there are ongoing efforts to develop CMOS chips that integrate photonics with electronics in large arrays for commercial protein sequencing.\cite{reed_real-time_2022} CMOS technology is also being developed for multi-analyte and multiplexed sensing of biomarkers by combining molecular and electronic components in a highly parallel architecture.\cite{fuller_molecular_2022}

While impressive, the commercialization success stories are few in comparison to demonstrations of novel sensors and techniques by academic groups and small companies.\cite{hu_super-resolution_2022, senevirathna_lab--cmos_2016, datta-chaudhuri_olfaction_2016,zeng_ultra-high_2021, le_quantum_2019, cho_high-resolution_2022, lin_machine_nodate, sajal_towards_2023,nizam_multimodal_2024,hall_256_2013} Academic groups must invest considerable resources to develop engineering capabilities and pay for mask sets used in custom CMOS design. Small companies, such as startups, face the additional burden of shouldering expensive licensing fees to access the electronic design automation (EDA) tools required to realize their designs. Furthermore, designing successful measurements and products often requires multiple iterations of design, fabrication, and testing, which cumulatively become cost-prohibitive. 

The preferred approach to reducing the investment needed to access custom CMOS designs is to participate in an MPW run, which allows multiple organizations to pool their resources to purchase a single mask set that contains all of their designs. The foundry fabricates wafers with all the projects from the MPW, dices them into small chips, each of which contains an individual design, and returns them to the participants. MPW runs have the additional advantage of minimizing the sharing of intellectual property between projects. Such a siloed framework is not well-suited for bioelectronics projects that require considerable postprocessing to integrate critical fluidic and organic elements. 

The small form factor ($\leq$1~cm per side) of chips returned from an MPW run makes them difficult to handle and increases processing challenges, which results in lower yield. To work around this limitation, the chips are mounted in larger carrier substrates. This is typically accomplished by adhering a handle wafer to the chip by using a photoresist\cite{liu_cell-lab_2004} or placing the chip face-down in a petri dish with epoxy and then curing the assembly.\cite{datta-chaudhuri_packaging_2014} While such approaches ease packaging challenges, the lack of precise alignment between the chip and substrate handle requires them to be processed using low-throughput techniques such as contact aligners or direct write lithography, resulting in lower yields than achievable when using high-throughput lithography tools. Finally, in cases where the handle wafers are too bulky or cumbersome to integrate with board-level electronics, chips are directly wire-bonded onto daughter boards with glob-top epoxy,\cite{gilpin_tracking_2022}, or by attaching plastic wells to create a fluidic reservoir.\cite{zeng_ultra-high_2021}

While the techniques discussed above allow fluidic integration, making electrical contact with the CMOS circuitry poses additional challenges. The chips must be passivated to minimize ion contamination of the CMOS layers, which prevents access to the top metal of the design, thereby hindering electrical contact.~\cite{delbruck_lessons_2020} Making robust electrical connections is further complicated by the surface roughness of the passivation layer and step height of the insulating surface, which follows the topography of the CMOS top metal layer. Finally, each individual integration step can take significant time, with minimal component reuse across projects. The net result is that there is a potential for small errors at each step that cumulatively cause significant delays and result in a low yield of functioning devices. 

In this work, we aim to address several challenges by utilizing the \textit{Nanotechnology Xccelerator}, a pilot multi-facility MPW program that employs open-source tools for generating CMOS designs. This initiative facilitates the efficient use of wafer-level back-end-of-the-line (BEOL) fabrication, enabling the development of prototypes that can be manufactured effectively. The rest of the manuscript is organized as follows. We first describe the \textit{Nanotechnology Xcellerator} (Sec.~\ref{sec:nanotechXccelerator}) followed by an example CMOS design that leverages this MPW program to design, fabricate, and integrate the components needed to measure DNA hybridization (Sec.~\ref{sec:design}). Next, we detail the instrumentation, data acquisition, and analysis infrastructure needed to efficiently acquire and process data for biological applications (Sec.~\ref{sec:instrumentation}). Finally, we demonstrate measurements of DNA hybridization using the system developed in this work (Sec.~\ref{sec:dna-measurements}).

\section{\label{sec:nanotechXccelerator}Nanotechnology Xccelerator: Manufacturable CMOS Prototyping}

The \textit{Nanotechnology Xccelerator }program, by design, has several unique features designed to promote nanotechnology research and prototyping that are amenable to developing manufacturable bioelectronics. The open-source nature of this program requires participants to collectively develop measurement test structures. This requirement allows groups to leverage design capabilities while lowering overall costs. Because all designs on the wafer are open-source, each participant can receive either individual flash fields (containing all projects) or entire wafers at a low cost to allow optimum post-processing for their final application.
    
The fabrication process is stopped after the last set of inter-layer vias are deposited and the surface planarized. This planarized surface with a well-characterized topology greatly simplifies BEOL processing. Additionally, the wafers received are not passivated, which allows a wide range of surface chemistries to be developed for different end applications. Furthermore, the planar substrate makes it straightforward to borrow integration strategies from traditional CMOS packaging, such as redistribution layers to fan out electrical contacts or integrate interposers to simplify packaging. Finally, the ability to work at the wafer level promotes the reuse of measurement instrumentation and infrastructure for multiple applications across MPW runs to enable affordable prototyping functions like a test vehicle that provides a low-parasitic interface to novel technologies.

\section{\label{sec:design}Wafer-Scale DNA Biochip Integration}

We offer a comprehensive guide to the design process, fabrication, and integration process of a CMOS device for measuring DNA hybridization using the wafer-level Nanotechnology Xccelerator MPW. Although certain aspects, such as the design of specific sensors, are single-use, we emphasize the modularity and component reuse enabled by working at the wafer scale throughout this section. This approach makes the tools developed here suitable for rapid and iterative prototyping while promoting manufacturability.

We first describe the CMOS design of a 25-unit-cell amplifier array to measure DNA hybridization and the results from its experimental validation (Sec.~\ref{sec:cmos-design}). This section is followed by a description of the BEOL fabrication of a redistribution layer to fan out the CMOS pads for easier electrical connections (Sec.~\ref{sec:beol-fab}). Notably, this fabrication step also provides a standardized interface that is available for other projects on the MPW tape-out. Finally, we describe a manufacturable integration approach that combines electrical and fluidic connections to allow measurements for biological applications (Sec.~\ref{sec:fluidics-design}). 


\subsection{\label{sec:cmos-design}CMOS Design and Validation}

\begin{figure*}
    \centering
    \includegraphics[width=1\linewidth]{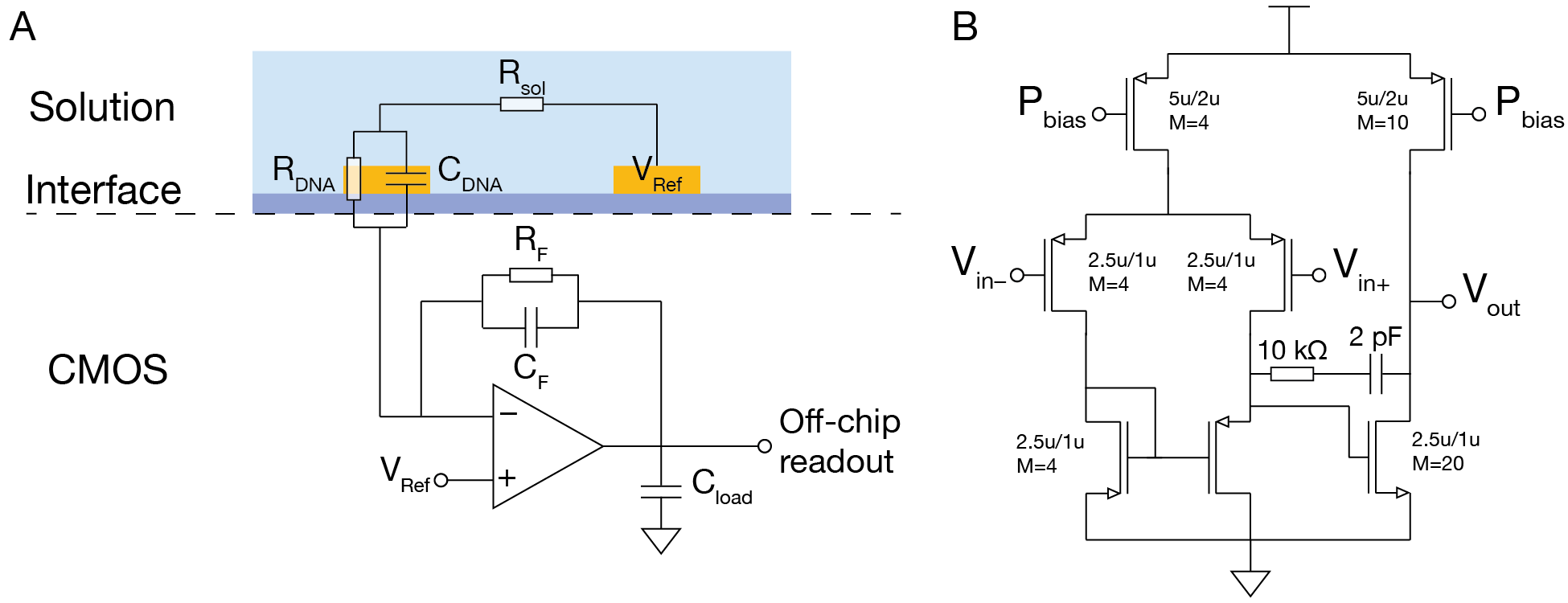}
    \caption{Model of the biological interface to CMOS and amplifier design. (A) A simplified schematic of the biological-electrical interface and embedded capacitive gain amplifier circuit. The feedback and compensation resistors and capacitors are implemented with high-resistance polysilicon and metal-insulator-metal capacitors. (B) A detailed circuit schematic of the amplifier, showing the differential input stage, the push-pull output stage, and the compensation resistor and capacitor.}
    \label{fig:cmos-design}
\end{figure*}

The CMOS circuitry used in this measurement is based on a specific physical model (see Fig.~\ref{fig:cmos-design}A) of the biological-electrical interface, which has been previously studied using electrochemical impedance spectroscopy (EIS).\cite{majikes_variable_2024, cho_high-resolution_2022} The circuit topology is chosen in order to be compatible with the experimental setup, which meaures the transfer function of the interface as a function of frequency, through a lock-in frequency sweep measurement. The circuit topology used is shown in  Fig.~\ref{fig:cmos-design}A, which shows a conceptual description of each sensing location, consisting of an amplifier connected to the working electrode (WE) and the reference electrode (RE). We modified each WE with ssDNA probes of a known sequence. When an analyte with a complementary sequence binds to the ssDNA probes, it changes the surface charge of the WE. To measure this change, we overlaid a DC bias with a small-amplitude AC voltage and applied it to the RE. The WE is connected to the negative terminal of an operational amplifier as shown in (Fig.~\ref{fig:cmos-design}A). Such a configuration operates as a capacitive gain amplifier in the designed frequency range, where the circuit's gain depends only on the ratio of the WE's capacitance($C_{DNA}$) to the feedback capacitance ($C_F$). This allows the circuit to convert changes in capacitance of the WE surface into a variable gain, amplifying small-signal input for frequency sweep measurements, enabling compatibility with previously developed measurement infrastructure\cite{le_optimal_2021}. 

The DNA-functionalized WE is described by a Randles circuit\cite{bard_allen_j_electrochemical_2000, majikes_variable_2024} that models the solution resistance ($R_{SOL}$) in series with the electroactive interface resistance ($R_{DNA}$) and capacitance ($C_{DNA}$) that are in parallel as seen from Fig.~\ref{fig:cmos-design}A. Our previous measurements that used a $\approx$1~mm diameter circular electrode provide baseline values for these circuit parameters ($R_{DNA}\approx1.5~M\Omega$, $C_{DNA}\approx38.7~nF$, and $R_{SOL}\approx918~\Omega$).\cite{majikes_variable_2024} In this design, the electrode sizes are considerably smaller ($\approx10~\mu m$ per side) and can be varied during the BEOL fabrication process to tune the measurement sensitivity. Assuming conventional area scaling for the electrode impedance, we expect an increase in $R_{DNA}$ to $\geq10~G\Omega$, and a decrease in $C_{DNA}$ to $\leq10~pF$, comparable to metal-insulator-metal (MIM) capacitors typically found in CMOS processes. 

In order to determine the frequency range at which capacitive sensing can be performed, we first model the transfer function of the circuit assuming an ideal amplifier. The input impedance of the Randles circuit is given by

\begin{equation}
    Z_{inp} = R_{sol} + \frac{R_{DNA}}{1+j{\omega}R_{DNA}C_{DNA}},
\end{equation}

where $j$ is the imaginary unit, and $\omega$ corresponds to the operating frequency. 
For $j{\omega}R_{DNA}C_{DNA}\gg1$, which is satisfied by $\omega >10~rad/s$ this expression simplifies to, 

\begin{equation}
\label{eq:input-impedance-simplified}
    Z_{inp} = R_{sol} + \frac{1}{j{\omega}C_{DNA}}.
\end{equation}

The impedance of the feedback circuit is then found to be, 

\begin{equation}
    Z_{f} = \frac{R_{f}}{1+j{\omega}R_{f}C_{f}},
\end{equation}

where $R_{f}$ and $C_{f}$ are the resistance and capacitance of the feedback path, respectively. For $j{\omega}R_{f}C_{f}\gg1$ this expression simplifies to 

\begin{equation}
\label{eq:feedback-impedance-simplified}
    Z_{f} = \frac{1}{j{\omega}C_{f}}.
\end{equation}

Note that the operating frequency range does not always satisfy the above condition, and hence, it sets the low-frequency cut-off for the circuit behavior. 

Assuming an ideal amplifier, we calculate the gain (G) by using the expression,

\begin{align}
    G = &-\frac{Z_{f}}{Z_{inp}} \\
       = &-\frac{C_{DNA}}{C_F}(\frac{1}{1+j{\omega}R_{sol}C_{DNA}}).
    \label{eq:open-loop-gain}
\end{align}

When $j{\omega}R_{sol}C_{DNA}\ll1$, which for $R_{SOL}\approx1~k\Omega$ and $C_{DNA}\approx10~pF$ is satisfied for $\omega <10^8~rad/s$ Eq.~\ref{eq:open-loop-gain} simplifies to,

\begin{equation}
    G = -\frac{C_{DNA}}{C_F}.
    \label{eq:gain}
\end{equation}

\begin{figure}
    \centering
    \includegraphics[width=1\linewidth]{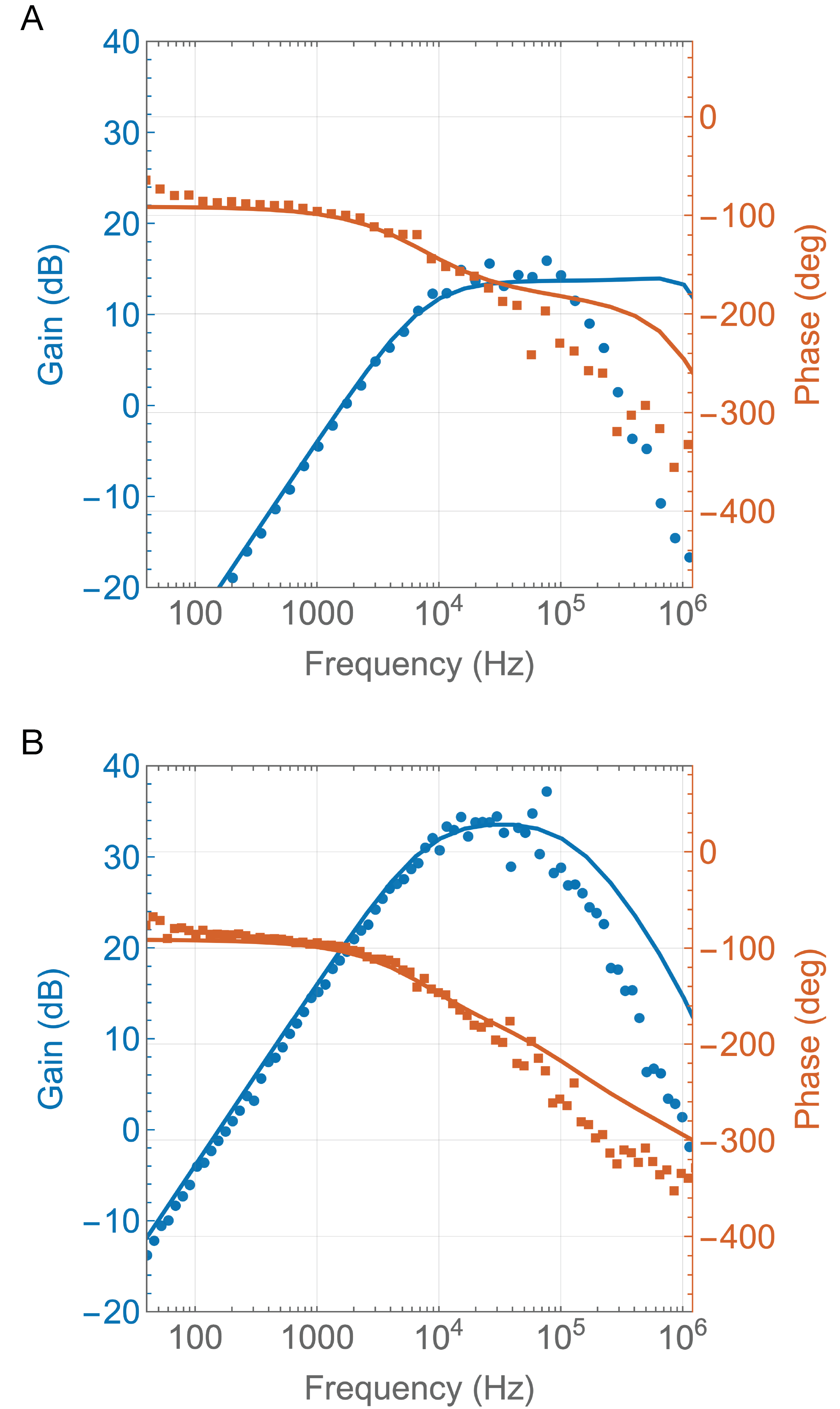}
    \caption{Gain and phase plots of capacitive gain amplifier. The solid lines show the theoretical circuit simulations, while the solid markers represent the experimentally verified circuit behavior. (A) Response of the amplifier when connected to a $10~pF$ input load. (B) Circuit response when connected to a $100~pF$ input load.}
    \label{fig:cmos-results}
\end{figure}

Hence, in an ideal system operated in the correct frequency range with a constant feedback capacitor, the circuit gain is directly proportional to the capacitance of the WE. In order for the circuit to have a lower cutoff frequency (as described in Eq~\ref{eq:feedback-impedance-simplified}) of $f\leq 10~kHz$, $R_F = 2~M\Omega$ and $C_F = 10~pF$ were chosen as parameters. Other parameters used in the design of the circuit are $R_{SOL} = 1~k\Omega$, $C_{DNA} = 10~pF$ or $100~pF$. 

In practical circuits, however, the amplifier is not ideal and must be designed under practical constraints such as gain, bandwidth, and drive strength. Typical operation parameters are variable gain from 5 to 50, based on input capacitance ($10~pF$ to $100~pF$), an operational frequency range of $10~kHz$ to $100~kHz$, and an ability to drive a capacitive load of $90~pF$ (approximately equal to 3 feet of coaxial cable). In order to meet these specifications, a two-stage amplifier was designed in a $130~nm$ process, with a high-gain differential input stage followed by a high-drive-strength push-pull output stage as shown in Fig.~\ref{fig:cmos-design}B. Because such an amplifier is not unconditionally stable, we added load compensation by using a resistor ($R_C=10~k\Omega$) and a capacitor ($C_C = 2~pF$).

The panels in Fig.~\ref {fig:cmos-results} show the theoretical and experimental gain and phase plots for the circuit shown in Fig.~\ref {fig:cmos-design}A with the above parameters for two input capacitance values, $10~pF$ and $100~pF$. The solid lines show the theoretical circuit simulations, while the solid markers represent the experimentally verified circuit behavior. We observed good agreement between the theory and the experiment for frequencies below $100~kHz$, after which the output signal began to distort and failed to follow the simulated behavior. From $100~Hz$ to $10~kHz$, we found that the circuit behavior followed the expected values defined by Eq.~\ref{eq:feedback-impedance-simplified}, for a frequency range where $j{\omega}R_{f}C_{f}\leq1$. This behavior flattens out at $f_L$ = $\omega_L/2\pi$ = $7.95~Hz$, which sets the lower limit of the valid frequency range in Eq.~\ref{eq:open-loop-gain}. The optimal operating range of the circuit, where the gain is maximum, was between $10~kHz$ and $100~kHz$. The realized bandwidth is adequate for practical measurements using these sensors, as described in the following sections. We employed a lock-in amplifier\cite{le_optimal_2021} that operates at a single frequency chosen from this range. This setup is effective for measuring the changes in capacitance that occur due to biomolecular binding.

\begin{figure}
    \centering
    \includegraphics[width=0.9\linewidth]{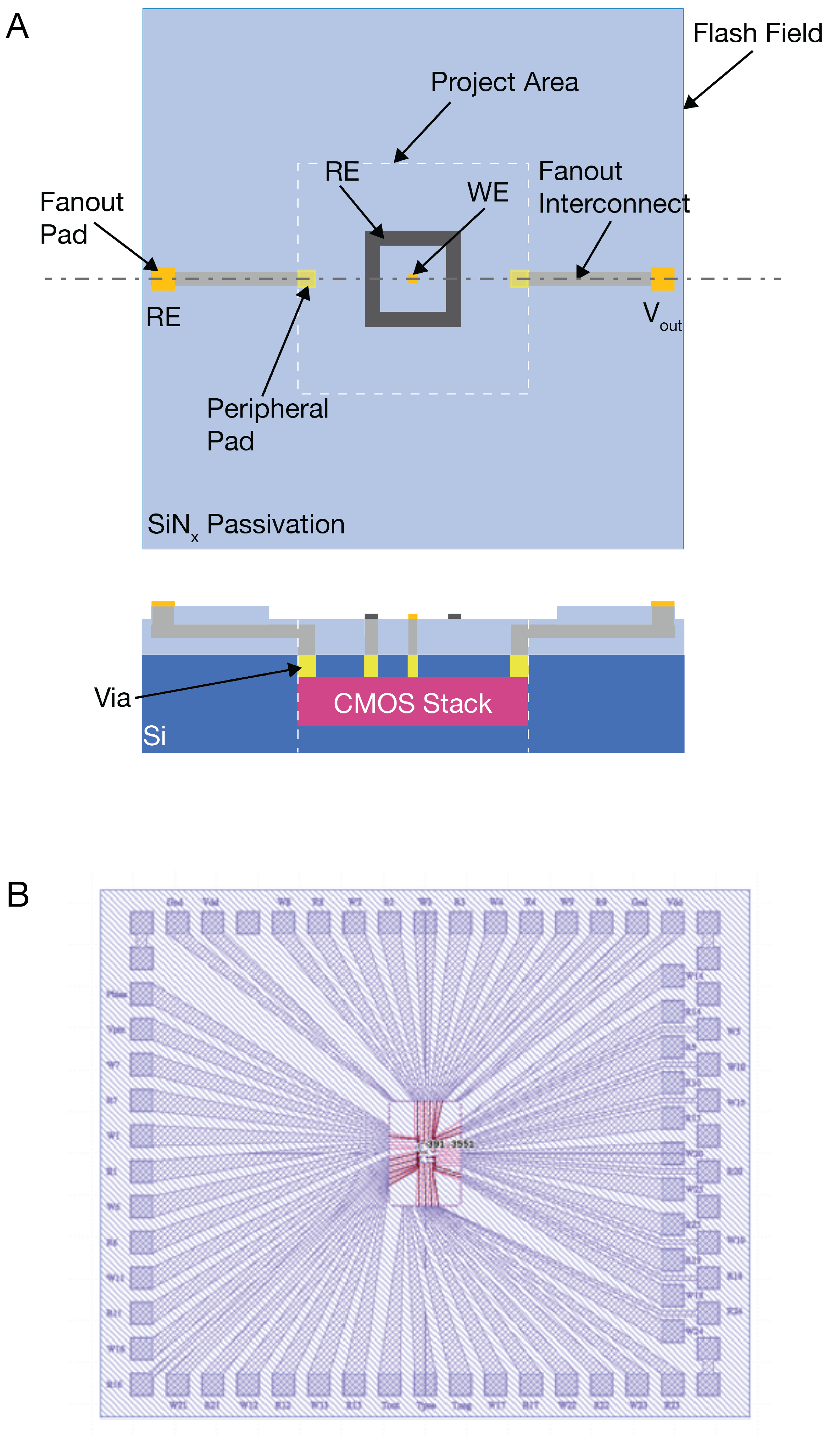}
    \caption{(A) Back end of line (BEOL) stack of CMOS chips. The schematic is not to scale. (B) The layout of fanout pads for the CMOS chips is shown in the central region, where the active area is located. The fanout interconnects extend over the other projects in the reticle and terminate in large pads measuring $\approx1~mm$ at the periphery. The overall chip size is $\approx31~mm \times 25~mm$.}
    \label{fig:CMOS-beol-fab}
\end{figure}

\subsection{\label{sec:beol-fab}Back End of Line Fabrication}

We utilized a BEOL process flow to create electrodes to access the underlying CMOS circuitry while passivating the chip from mobile ion contamination during measurements. The biggest challenge with adding metal and passivation layers to foundry-fabricated CMOS wafers is adhering to a strict thermal budget. The wafers cannot be exposed to temperatures exceeding $450~^{\circ}C$ without initiating a diffusion of dopants that cause damage to the chips.\cite{takeuchi_thermal_2005} Therefore, we limited all processes to a maximum temperature of $300~^{\circ}C$. We received $200~mm$ wafers from the foundry with vias in place for the final metal layer (metal 4 in all chips discussed here). The foundry stopped processing after this step, returning planar wafers with a surface roughness of $\approx1~nm$ and maximum step height of $\approx10~nm$. This attribute of the \textit{Nanotechnology Xcellerator} MPW allows us to use the same signal redistribution strategy in multiple wafer projects, allowing for a common electrical interface and greater reusability for different applications.

The process flow for the mid-process wafers is shown in Fig.~\ref{fig:CMOS-beol-fab}A. 

\begin{enumerate}[(i)]
    \item We first cored $100~mm$ diameter wafers from the $200~mm$ substrates received from the foundry to facilitate easier processing within our cleanroom.
    
    \item We covered the vias with $\approx100~nm$ thick titanium (Ti) pads that were defined by using optical lithography and a metal deposition and lift-off process. This step was performed to protect the vias and also provide an etch stop for future processing steps.
    
    \item\label{item:beol-nitride-window} We followed this step by depositing $\approx1~\mu m$  of low-stress Silicon Nitride by using plasma-enhanced chemical vapor deposition (PECVD). The PECVD nitride served two purposes: it provided electrical passivation for the fanout pads described in subsequent steps and also prevented mobile ion contamination of the underlying CMOS layers from Potassium and Sodium ions present in biological solutions. We then lithographically patterned windows and etched back the nitride films up to the Ti pads by using reactive ion etching (RIE) with a CF$_4$ chemistry.
    
    \item\label{item:beol-Ti-backfill} The windows were then back-filled with $\approx1.1~\mu m$ of Ti by using an additional lithography step followed by metal deposition and lift-off.
    
    \item Leads to fan out the peripheral pads around the CMOS active area (Fig. \ref{fig:CMOS-beol-fab}B) were fabricated by using an additional lithography step followed by the deposition of $\approx1~\mu m$ of Ti followed by lift-off. A critical feature of our MPW that allows fan-out is the fact that the leads can be routed on top of the other projects in the reticle, resulting in a large handle area of $\approx31~mm\times25~mm$. Furthermore, the nitride layers that were deposited in the previous step electrically isolate the unused projects, thereby minimizing failures.
    
    \item Following the definition of the fan-out leads, we deposited $\approx1~\mu m$ of Silicon Nitride by using PECVD followed by the definition of windows and backfilling with Ti using the process outlined in steps \ref{item:beol-nitride-window} and \ref{item:beol-Ti-backfill}. 
    
    \item  Next, the contact pads, working electrodes, and dice lines were patterned by using optical lithography, followed by the deposition of $\approx10~nm$ of Ti for an adhesion layer and $\approx100~nm$ of gold (Au), and lift-off. Finally, the reference electrodes were patterned by using optical lithography, followed by the deposition of $\approx10~nm$ of Ti for an adhesion layer and $\approx100~nm$ of platinum (Pt), and lift-off. 
    
    \item The finished wafers were then diced into chips for assembly into the fluidic cell described in Sec. \ref{sec:fluidics-design}. 
\end{enumerate}

\begin{figure}
    \centering
    \includegraphics[width=0.9\linewidth]{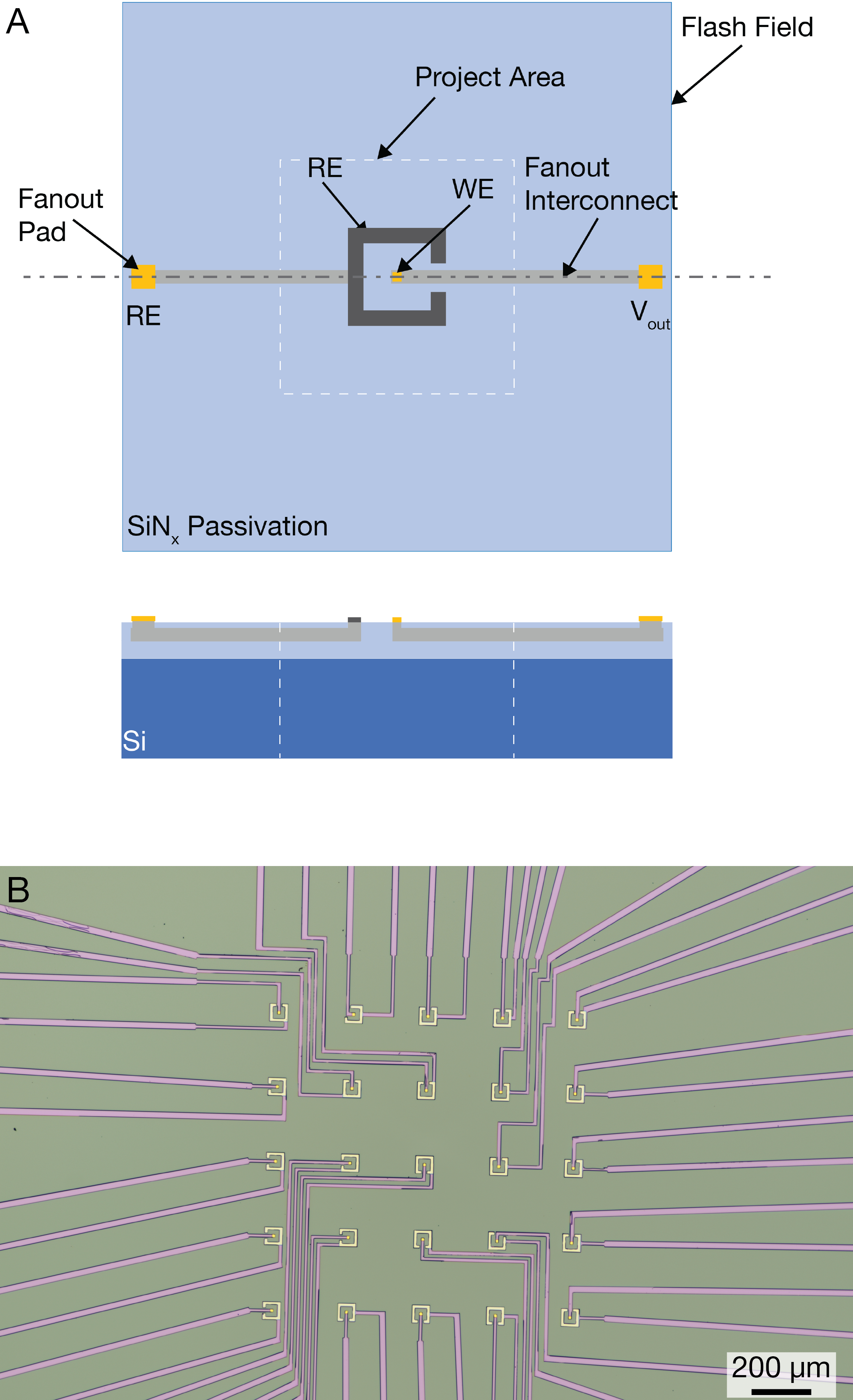}
    \caption{(A) Back end of line (BEOL) process stack for the test chips.  The schematic is not to scale. (B) Enlarged image of the working and reference electrodes at the center of the test chips. Fanout interconnects from the working and reference electrodes are shown in pink.}
    \label{fig:Test-beol-fab}
\end{figure}

Since the lead times for CMOS fabrication can be substantial, on the order of months, we designed and fabricated test chips that lacked CMOS circuitry to develop and validate packaging and integration strategies. We designed the test chips, shown in Fig.~\ref{fig:Test-beol-fab}B, to have an identical pad layout as the CMOS chips. To account for the lack of circuitry, additional leads were added to route output signals to the peripheral contact pads so they could be measured using external instrumentation. The fabrication steps for these test chips (Fig.~\ref{fig:Test-beol-fab}A) were as follows: 

\begin{enumerate}[(i)]
    \item First, $\approx2.8~\mu m$ of thermal oxide was grown on the 100~mm wafers using a wet oxidation process. The leads and contact pads were deposited \textit{via} a lithography step, followed by deposition of $\approx100~nm$ Ti and then $\approx90~nm$ Pt.
    
    \item  Next, $\approx200~nm$ of low-stress silicon nitride was deposited using a PECVD process. The windows were patterned lithographically for the reference electrodes and then etched into the nitride layer by using an RIE process.
    
    \item Finally, the dice lines and windows for the peripheral contact pads and the working electrodes were lithographically patterned and then etched using an RIE process. Once etched, $\approx100~nm$ of Ti and $\approx100~nm$ of Au were deposited by using an e-beam evaporation process, followed by lift-off. Metal deposition was designed to ensure that the electrodes rise up to or above the silicon nitride layer $\approx200~nm$. The wafers were diced into individual chips as shown in Fig.~\ref{fig:Test-beol-fab}B.
\end{enumerate}



\subsection{\label{sec:fluidics-design}Packaging Design and Integration}

The ability to prototype at the wafer scale and to use BEOL techniques to redistribute contact pads, as described in Sec.~\ref{sec:beol-fab}, eases packaging challenges and can allow solutions that are (i) modular to allow components such as fluidics inserts to be rapidly exchanged, (ii) manufacturable because they use standard industrial processes such as CNC machining and 3D printing, and (iii) flexible to provide standard interfaces that allow the package to be configured for chips built for different applications.


\begin{figure}
    \centering
    \includegraphics[width=0.8\linewidth]{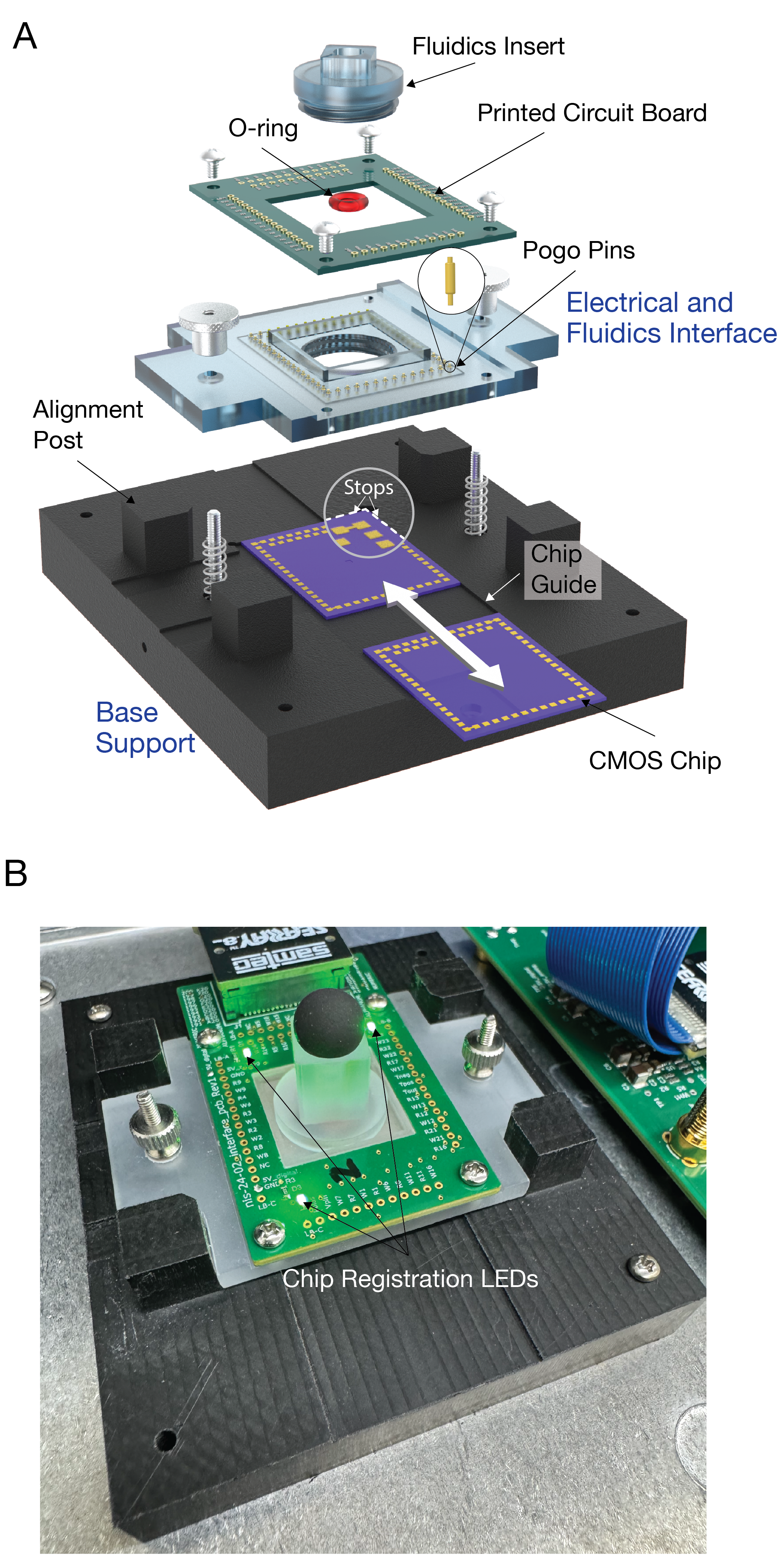}
    \caption{(A) CAD model of the package that integrates mechanical, fluidic, and electrical components. Channel guides and stops ensure positive registration of the chip, while the upper mechanical component supports the electrical interface and interchangeable fluidic inserts. (B) A photograph of the assembled fluidic package. The printed circuit board (PCB) implements the electrical interface and also validates chip registration and alignment through the use of three light-emitting diodes (LEDs) that illuminate when the pogo pins contact the chip pads.}
    \label{fig:package-design}
\end{figure}

Our packaging solution, which strives to achieve the above attributes, is shown in Fig.~\ref{fig:package-design}. The package comprises two primary structural elements, the \textit{base support} that houses and aligns the chip, and the \textit{electrical and fluidics interface} that provides structural support for electrical and fluid connections. In the example shown in the figure, the \textit{base support} was CNC machined while the \textit{electrical fluidics interface} was 3D-printed with a biocompatible resin, although it is also possible to fabricate these components using CNC machining for enhanced mechanical durability or chemical resistance. Both components are discussed in detail next.

\textit{\textbf{Base Support:}}
This element simplifies the alignment of the chip to ensure reproducible contact with the \textit{electrical and fluidics interface}. Guides constrain the chip and help alignment at the stop (\textit{dashed white lines}) at the top right corner of the \textit{base support} as seen in Fig.~\ref{fig:package-design}A. The thin channels to the left allow the chip to be pushed against the guide to ensure positive registration and alignment. When measurements are complete, we ejected the chip by using a pin guided by a channel on the back of the \textit{base support}. The movement of the \textit{electrical and fluidics interface} when lowered to make electrical contact with the chip is constrained by four alignment posts integrated into the \textit{base support}. Finally, the \textit{base support} houses small spring-loaded bolts that are compressed by thumb screws that control the engagement of the pogo pins with contact pads on the chip. 

Fig.~\ref{fig:package-design}B shows a version of the \textit{base support} used in this study that was 3D-printed using the thermoplastic Polyoxymethylene. We designed the parts to be easily adapted for injection molding for low-cost, high-volume applications or to allow tighter mechanical tolerance using CNC machining.

\textit{\textbf{Electrical and Fluidics Interface:}}
The \textit{electrical and fluidics interface} (Fig.~\ref{fig:package-design}A) supports the critical functions of making electrical contact with the chip to apply control signals or read data and delivering fluids \textit{via} a customizable insert. The redistribution layer described in Sec.~\ref{sec:beol-fab} simplifies the electrical interface by allowing contact pads that are $\approx1~mm$ on an edge with an inter-pad pitch of $\approx1.7~mm$. The large pad size makes the design tolerant to angular alignment errors of up to $\approx \pm1^{\circ}$. Our pad layout also permits using spring-loaded pogo pins (Fig.~\ref{fig:package-design}) to create electrical connections. The electrical and fluidic interface body contains double-ended pogo pins in pockets sandwiched between the printed circuit board (PCB) and the chip. This design eliminates the need for wire bonding and soldering, allowing for rapid chip exchange. 

We verified the proper insertion of the CMOS chip and its electrical contact with the \textit{electrical and fluidics interface} by designing a loopback interface comprised of pairs of pads on three corners of the chip as seen from the magnified region (\textit{top right}) of the CMOS chip in Fig.\ref{fig:package-design}A. We connected one pad in the pair to the common ground of the circuit. The second pad was connected to a light-emitting diode (LED), which was powered by a $5~V$ power supply \textit{via} a current-limiting resistor. When both pogo pins make positive contact with their corresponding pads, the LED illuminates as seen in Fig.~\ref{fig:package-design}B. The illumination of all three LEDs indicates positive registration against the chip guides, the chip is oriented properly, and there are no angular misalignments beyond the design tolerance.

In addition to electrical connections, we developed modular fluidics inserts to facilitate sample delivery, as seen in Fig.~\ref{fig:package-design}. The fluidic inserts can be threaded from the top of the \textit{electrical and fluidics interface}. In the current study, an open well insert with a volume of $\approx160~\mu L$ was used to allow rapid fluid exchange using standard laboratory pipettes. Importantly, this insert can be customized for specific applications, for example, by replacing the open well insert with one that incorporates microfluidic channels that allow rapid fluid exchange. 

\section{\label{sec:instrumentation}Instrumentation}

\begin{figure}
    \centering
    \includegraphics[width=1\linewidth]{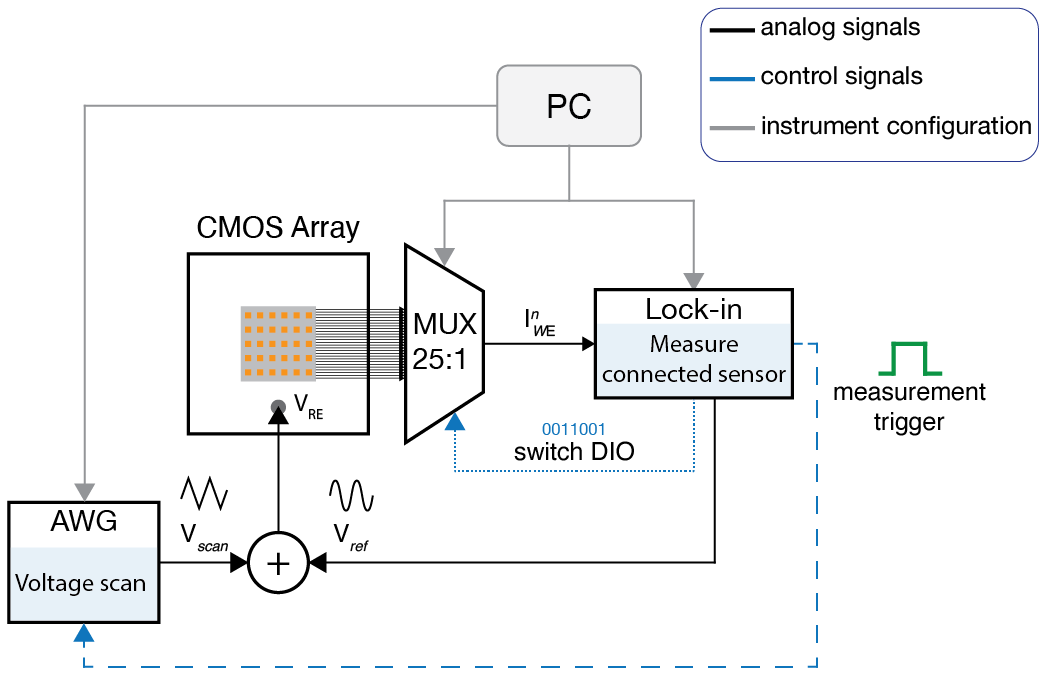}
    \caption{Schematic of the hardware setup for multiplexed sensing. The on-board multiplexers are controlled by using a digital IO (DIO) interface. The measurements are initiated by using a trigger signal from the lock-in amplifier, which starts the voltage scan using the arbitrary waveform generator (AWG) and also arms the measurement for each sensor as described in the main text.}
    \label{fig:instrumentation-layout}
\end{figure}

\subsection{Instrumentation Layout and Signal Timing}

We performed the measurements by using a two-electrode electrochemical cell with a single global RE and 25 independently addressed WEs. Instead of individual REs paired to each WE as designed, we employed a global Ag/AgCl quasi-reference electrode in this study. We then sequentially connected each WE in the array to a lock-in amplifier (LIA) \textit{via} a 25:1 multiplexer (MUX) as illustrated in Fig.~\ref{fig:instrumentation-layout}. The measurements were initiated with the transmission of a global \textit{measurement trigger} (indicated in \textit{green}). At the same time, we connected the first sensor to the LIA by using the digital I/O (DIO) interface through the MUX. After completing the measurements with the connected electrode, we sequentially connected the subsequent sensors to the LIA according to a scan list.

The global \textit{measurement trigger} also initiated a voltage scan, $V_{scan}$, controlled by the arbitrary waveform generator (AWG) and summed with the LIA reference AC signal. The \textit{switch pulse} is then overlaid on top of $V_{scan}$. While we demonstrate this ability with a triangle wave, it is straightforward to adapt this approach for other waveforms, such as a square wave or even a constant DC voltage. This technique allows us to supply arbitrary waveforms to interrogate the samples that can be tailored to particular applications. Multiple scans of the array allow us to robustly sample $V_{scan}$ each sensor, measuring a slightly different region of the waveform. By saving both the measured signal and the corresponding segment of $V_{scan}$, we can accurately reconstruct the signal-voltage response for the entire array. Finally, because the phase relationship between the scan pulse and $V_{scan}$ is well-defined and adjustable, we are able to repeat the measurements with a different starting phase for $V_{scan}$ to improve coverage of the desired waveform and tune the granularity of the signal-voltage data.

\subsection{\label{sec:data-acq}Data Acquisition}

We used purpose-built Python scripts\cite{national_institute_of_standards_and_technology_pylabauto_2024} run on a PC (Fig.~\ref{fig:instrumentation-layout}, \textit{gray lines}) to handle instrument configuration and data handling. The scripts abstracted communications with the instrumentation either over the universal serial bus (USB) or the general-purpose interface bus (GPIB) for seamless operations. The configuration included initializing the MUX with the scan list that enumerated the connection order of the measurement. We configured AWG with the waveform for $V_{scan}$ that was summed with the lock-in reference signal. We initiated the measurements by programming the LIA to emit the \textit{measurement trigger} signal as seen in Fig.~\ref{fig:instrumentation-layout}. Measurements of each connected WE were saved separately to disk to simplify data analysis as described in the next section.  

\begin{figure}
    \centering
    \includegraphics[width=1\linewidth]{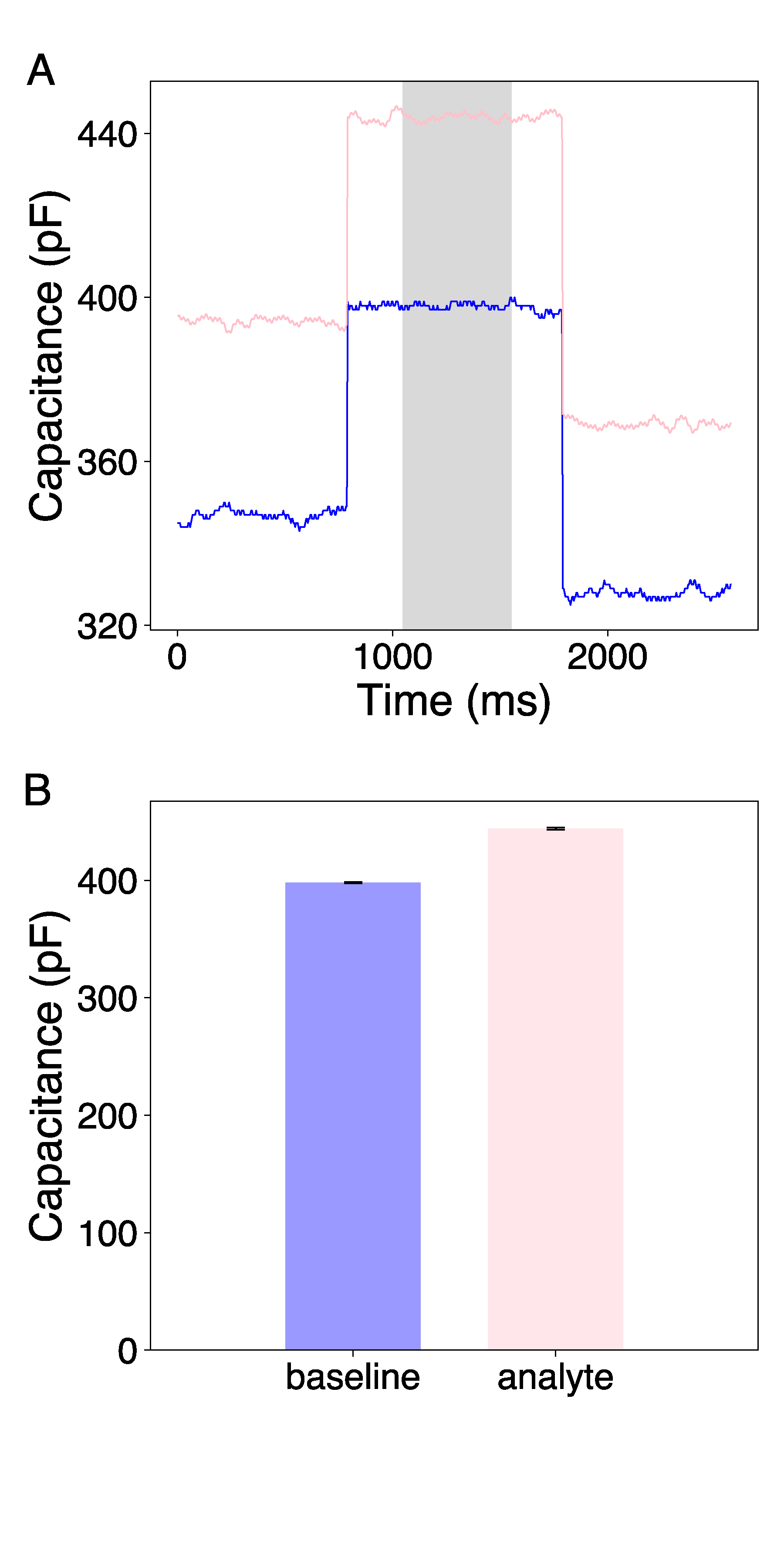}
    \caption{Representative data from a single sensor. (A) Capacitance response for baseline single-stranded DNA (\textit{blue}) and hybridized double-stranded DNA (\textit{pink}). The measurements were performed under an applied AC field with amplitude $V_{AC, pk}=50~mV$ and frequency of 500~Hz. (B) The capacitance of the single-stranded DNA (\textit{blue}) and double-stranded DNA (\textit{pink}) was estimated by averaging the steady-state signal across the shaded region shown in panel A, which represents $\approx50~\%$ of the dwell time. The error bars represent one standard deviation.}
    \label{fig:data-analysis}
\end{figure}

\section{\label{sec:dna-measurements}DNA Measurements}

\textit{\textbf{DNA Sequences and Preparation:}}
The ssDNA anchor sequence used was /5ThioMC6-D/ TAGTCGTAAGCTGATATGGCTGATTAGTCGGAAGCATCGAACGCTGAT, and the analyte sequence was ATCAGCGTTCGATGCTTCCGACTAATCAGCCATATCAGCTTACGACTA. The control strand sequence was TTTATACAGGAGCAATCATAAGGG CATTAGACCCAGCCC. We purchased both DNA sequences in $100~\mu mol/L$ ($\mu M$)  in 1 x TE buffer containing $10 ~mmol/L$ ($mM$) Tris and $100~\mu mol/L$ ($\mu M$) Ethylenediaminetetraacetic acid (EDTA). 1x TAE buffer comprising $\approx40~mmol/L$~($mM$) Tris base, $\approx20 ~mmol/L$ ($mM$) acetic acid, $\approx1~mmol/L$~($mM$) EDTA, was diluted from commercially available 50x TAE buffer with deionized water. 

To prepare the ssDNA anchor, $10~\mu L$ of $400~nmol/L$ ($nM$)  ssDNA was reduced with $30~\mu L$ of tris(2-carboxyethyl)phosphine (TCEP) as the reducing agent for 2 hours. Similarly, $10~\mu L$ of 6-mercapto-1-hexanol (MCH) was reduced with $30~\mu L$ of TCEP for 2 hours. Following reduction, the mixtures were centrifuged at 1000 relative centrifugal force (RCF), and the supernatant solution was collected for subsequent experiments.

\textbf{\textit{Electrode Functionalization and Electrical Measurements:}} 
The chips with an array of 25 gold electrodes were prepared for measurements by first cleaning and modifying them with thiol chemistry. After the chamber was pre-cleaned for 30 minutes, the chips were cleaned for 30 minutes in a UV Ozone cleaner. The cleaned electrodes were immediately immersed in $50~mmol/L (mM)$ sulfuric acid solution, and each electrode was chemically cleaned by scanning the applied potential from $-0.25~V$ to $+1.5~V$ relative to an Ag/AgCl reference electrode for at least 10 cycles at a sweep rate of $0.5~V/s$. 

The chips were then rinsed five times with DIW and immediately immersed in a solution of $\approx10~\mu mol/L$ ($\mu M$) of the ssDNA anchor in TAEM buffer for $\approx$1~hour at room temperature. This functionalization with an ssDNA anchor strand was followed by a second incubation step to passivate the unreacted gold electrode surface by immersing the chips in a solution comprised of $\approx1~mmol/L$~($mM$) of MCH in ethanol for $\approx16~hours$. Finally, the chips were thoroughly washed with running buffer solution (RBS) at least five times before being used for electronic measurements. For these measurements, RBS was prepared with 1x TAE buffer containing $50~mmol/L$ ($mM$) NaCl, $12.5~mmol/L$ ($mM$) MgCl$_2$, $200~\mu mol/L$ ($\mu M$) K$_4$Fe(CN)$_6$, and $200 ~\mu mol/L$ ($\mu M$) K$_3$Fe(CN)$_6$. 

All electronic measurements were performed by using a two-electrode electrochemical cell with a $\approx 160~\mu$L sample volume. Capacitance measurements were performed at a fixed frequency of $500~Hz$ with an applied AC voltage with amplitude, $\approx50~mV_{PP}$ and a constant DC offset of $\approx0~V$ relative to an Ag/AgCl reference electrode and analyzed as shown in Fig.~\ref{fig:data-analysis}. 

\textbf{\textit{DNA Hybridization:}}
Baseline measurements of the gold electrodes with ssDNA anchor and MCH were performed under RBS. Strong chemical interactions exist between the DNA bases and the gold surface due to $\pi$-metal and van der Waals forces. These interactions can cause immobilized single-stranded DNA to adsorb nonspecifically onto the gold electrode, hindering hybridization and signal generation. To mitigate this behavior, we washed the cell five times with RBS after first incubating it with the DNA analyte strand that was dissovled in RBS at a concentration of $1~mmol/L$ ($mM$) for 60 minutes. We then measured the DNA hybridization two hours after the rinse step to allow the DNA to adopt a more upright conformation, improving measurement reproducibility.

\begin{figure}
    \centering
    \includegraphics[width=1\linewidth]{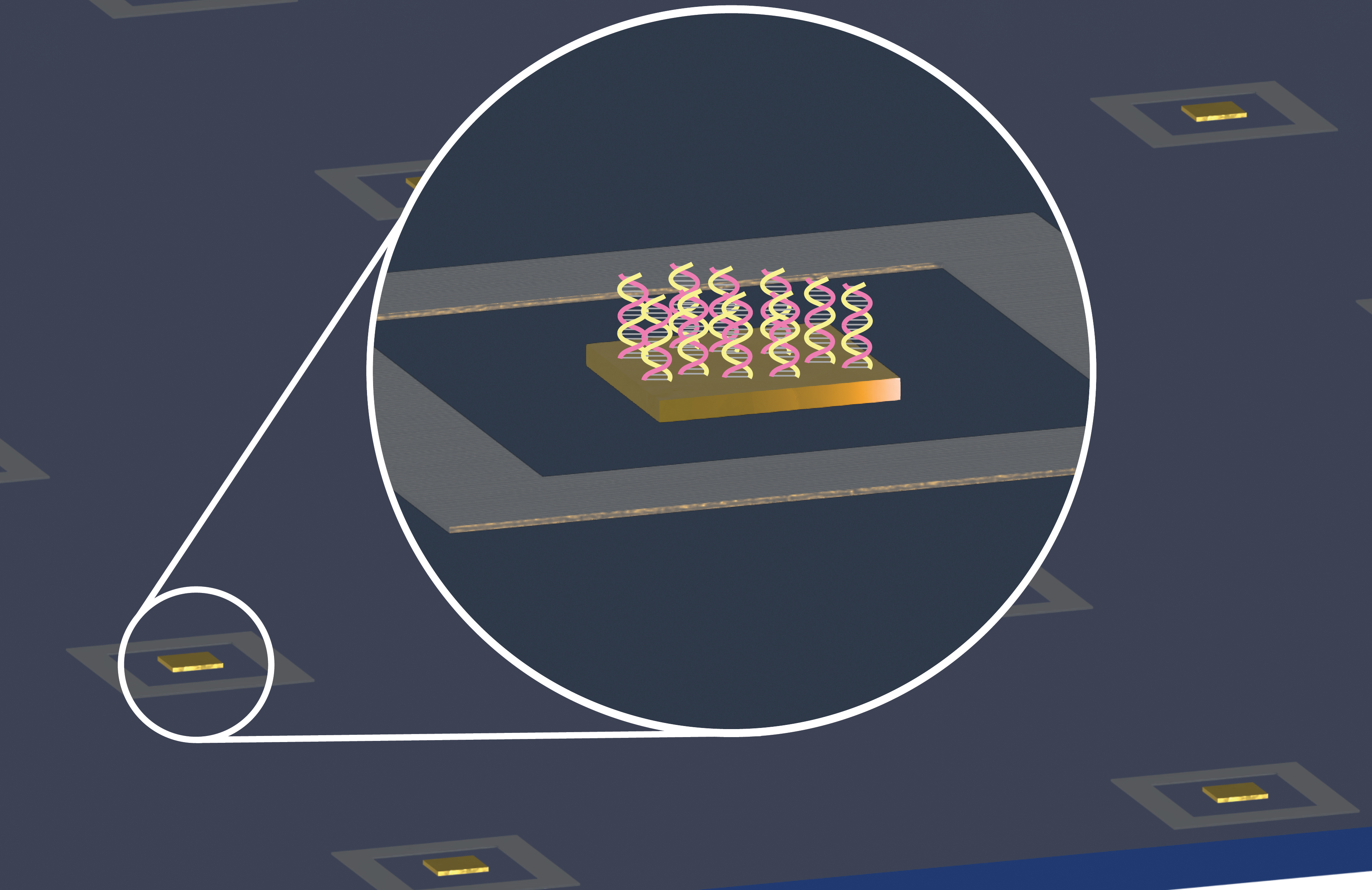}
    \caption{DNA chip measurement schematic. The square-shaped gold sheet represents one of the 25 ($\ approx$10~$\mu$m per side) gold electrode arrays. DNA anchors (\textit{magenta}) are attached to the surface using thiol chemistry. The analyte (\textit{yellow}) was added to bind the DNA anchors through complementary sequence hybridization.
}
    \label{fig:dna-schematic}
\end{figure}

\section{\label{sec:dna-results}DNA Hybridization Results}

\begin{figure*}
    \centering
    \includegraphics[width=0.8\linewidth]{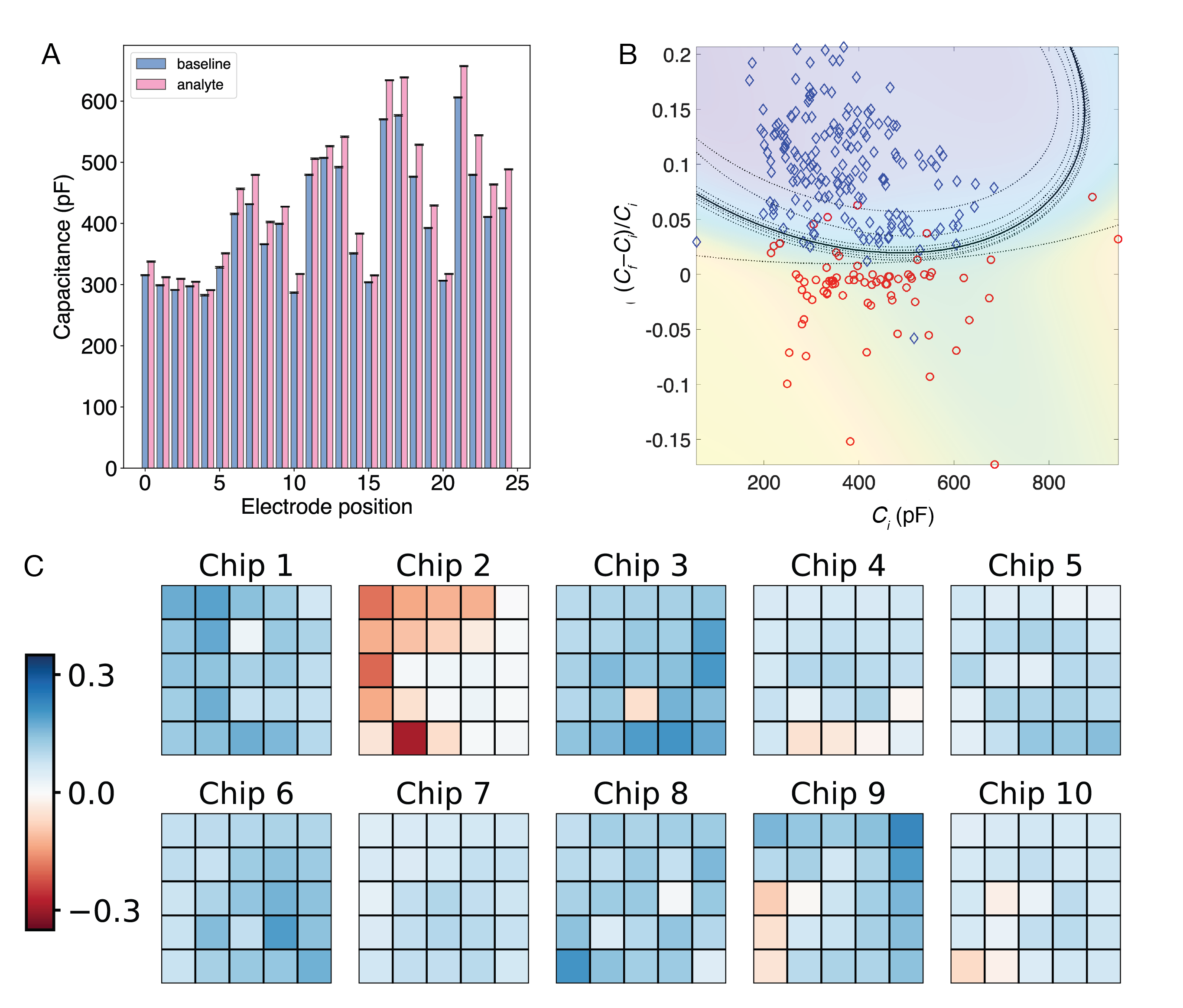}
    \caption{A. Representative capacitance measurements from all 25 electrodes on a single chip under baseline conditions (ssDNA) and after hybridization with an analyte with a complementary sequence (dsDNA). B. Classification analysis characterizing the performance of ten chips comprised of 25 electrodes each.  Blue diamonds are measurements from individual sensors exposed to the analyte, whereas red circles are control sensors that were not exposed to the analyte. The colored surface and contours quantify the relative likelihood $R(x)$ that a sensor with a given measurement value $x$ is being bound by the analyte.  From bottom to top, the contours correspond to values $R(x)=0.11,0.25,0.43,0.67,1.0,1.5,2.3,4.0,9.0$. C. Heatmaps showing the spatial distribution of relative capacitance changes across ten measured chips, with electrode positions corresponding to their actual physical locations.}
    \label{fig:dna-measurements}
\end{figure*}

We measured the interfacial capacitance of gold electrodes functionalized with single-stranded DNA (ssDNA) anchors in the presence of an applied electric field to detect the specific hybridization of DNA sequences that have a complementary sequence to the probe strands attached to the electrode surface, as shown in Fig.~\ref{fig:dna-schematic}. When an electric field is applied, it attracts counter-ions from the bulk solution to the surface, resulting in the formation of an electrical double layer. This double layer behaves like a nanoscale capacitor and is influenced by the ssDNA, yielding the baseline measurements shown in Fig.~\ref{fig:data-analysis}. 

The binding of complementary DNA strands (analytes) to the anchored ssDNA alters the structure of this double layer, resulting in a change in interfacial capacitance from the baseline values, also demonstrated in Fig.~\ref{fig:data-analysis}. The increase in the capacitance, relative to the baseline value, can be understood from the change in conformation adopted by DNA upon hybridization. Surface-bound ssDNA strands typically lie flat against the gold surface, obstructing parts of the electrode from interacting directly with the electrolyte. However, upon hybridization, the double-stranded DNA adopts a more rigid and upright orientation, exposing additional electrode surface area that was previously hidden by the surface-adsorbed ssDNA and resulting in an overall higher capacitance.

Capacitance measurements, at a constant potential $0.0~V$ relative to an Ag/AgCl reference electrode, shown in Fig.~\ref{fig:dna-measurements}, were recorded for both the baseline ssDNA case and after the addition of the analyte with a complementary sequence. We measured a total of ten chips, each with 25 electrodes. Following the addition of the analyte, we observed an increase in the capacitance, indicating successful detection of hybridization between the analyte strand and the ssDNA anchor (Fig.~\ref{fig:dna-measurements} B).  Across ten chips, $n=250$ electrodes in total, DNA hybridization resulted in an average capacitance change of $(24.5 \pm 3.9)~pF$ or a relative change of $(7.8\pm 0.9)~\%$ ($k=2, \approx95~\%$ confidence) upon hybridization. 

To verify the specificity of the sensors, a non-complementary DNA strand was used as a negative control and incubated under the same conditions as the complementary analyte strands. The probability of hybridization with the non-complementary strand was estimated using NUPACK simulations, which were conducted at a temperature of 25°C and a concentration of $10~nmol/L (nM)$, allowing for the formation of complexes with up to four strands. Under these conditions, NUPACK predicted a binding probability of less than 1~\% for the non-complementary control strand interacting with the ssDNA anchor. For these control electrodes ($n=75$), the average change in capacitance after incubating with the non-complementary sequence was measured to be $(-5.5 \pm 4.8)~pF$  or a relative change of $(-1.3\pm 0.9)~\%$ ($k=2, \approx95~\%$ confidence). This result confirms that non-specific binding did not produce a significant change in the measured capacitance.

To assess the performance of the biosensors in accurately detecting the analyte, we performed a homotopy-classification analysis\cite{patrone_analysis_2024, patrone_inequalities_2025, patrone_probabilistic_2025} as shown in Fig.~\ref{fig:dna-measurements}B.  The central idea of this analysis is to first assume that ``positive'' sensors (i.e., those exposed to analyte) yield measurements $x$ according to some distribution $P(x)$, whereas control sensors yield measurements $x$ with a different distribution $N(x)$.  In this work, $x$ is taken to be a two-dimensional vector $x=(x_1,x_2)$ with components,

\begin{align}
x_1 &= C_i \\
x_2 &= \frac{C_f - C_i}{C_i},
\end{align}

where $C_i$ is the initial capacitance, and $C_f$ is the capacitance after exposure to the analyte or control sample. We then construct a family of classification boundaries that correspond to different values of the relative likelihood $R(x)=N(x)/P(x)$.  Physically, we can interpret $R(x)$ as quantifying the extent to which a positive sensor would generate a measurement $x$ as compared to a negative sensor.  

To quantify the probability $\Pr[+|r]$ that a given sensor detects an analyte, it is necessary to first estimate the prevalence $q$ (or fraction) of positive sensors.  This task, which has been addressed in a number of fields, has a rich history that is beyond the scope of the current manuscript (see Ref.~\onlinecite{patrone_probabilistic_2025} and the references therein for more details).  Here we  note that given $q$, the probability $\Pr[+|r]$ is given by

\begin{align}
\Pr[+|r] = \frac{q}{q+(1-q)R(x)}.
\end{align}

Absent a reasonable estimate of $q$, the $R(x)=1$ boundary (bold contour in Fig.~\ref{fig:dna-measurements}B) can also be used as a classifier for determining when a sensor is positive.  For the data presented here, the characteristic accuracy of this classifier is at least $95~\%$ for any value of prevalence. 

The visualization of the electrode arrays as 5×5 heatmaps, shown in Fig. \ref{fig:dna-measurements}C, facilitates a quick assessment of both inter-chip and intra-chip variation. The figure reveals that nine out of the ten measured chips exhibited consistent performance, predominantly showing positive capacitance changes, with relatively few electrodes displaying negative values. Only one chip demonstrated behavior in which a significant portion of the electrodes had either negative or near-zero capacitance changes following incubation with the analyte. This anomaly may be attributed to localized surface contamination, fabrication defects, or handling errors specific to that chip. When excluding this outlier, 95~\% of the electrodes on the remaining chips showed a relative capacitance change ranging from 2~\% to 21~\%.

\section{Conclusions}
We present a manufacturable method for creating CMOS sensors suitable for healthcare by addressing key research areas that currently hinder innovation. A significant advancement supporting our approach is the Nanotechnology Xcellerator program, an open-source MPW initiative that enables all participants to reuse designs while promoting research in BEOL fabrication techniques to yield innovative devices. The program's flexibility and the ability to prototype using full wafers are distinguishing features that significantly enhance the manufacturability and yield of biosensors. Moreover, these advantages are achieved while benefiting from the cost-effectiveness associated with participating in MPW tapeouts.

The ability to work with larger silicon substrates alleviates packaging constraints. The larger substrates allowed us to utilize standard packaging strategies, such as fan-out techniques, to ensure repeatable electrical connections, thereby further improving yield. Our approach also facilitated the modular integration of mechanical, electrical, and fluidic components, resulting in a comprehensive packaging solution. Additionally, our packaging was developed using either 3D-printed or machined parts to ensure manufacturability.

Finally, we leveraged the advances in device processing and packaging to demonstrate a biosensor array that was used to measure the hybridization of DNA strands with a complementary sequence. We demonstrated the performance of the sensors by using data from ten individual chips (a total of 250 sensors) and three control chips (with 75 sensors) that were incubated with a non-complementary sequence. The data allowed us to perform homotopy classification, which allowed us to quantify the likelihood of the sensors accurately detecting the analyte. This analysis showed that the characteristic accuracy of identifying the analyte was at least 95~\%.

\begin{acknowledgments}
This research was performed, in part, at the NIST Center for Nanoscale Science and Technology, nanofabrication facility.
\end{acknowledgments}



\section*{References}
\bibliography{references}

\end{document}